\documentstyle[fleqn,12pt,psfig]{article}
\input{epsf.tex}
\newcommand{\be}{\begin{equation}}
\newcommand{\ee}{\end{equation}}
\newcommand{\bea}{\begin{eqnarray}}
\newcommand{\eea}{\end{eqnarray}}
\newcommand{\gv }{g_{_V}}
\newcommand{\ga }{g_{_A}}
\newcommand{\ph }{P_{_H}}

\begin{document}
~\\
{\large \sc  Energy Dependence of Solar Neutrino Suppression and Bounds on the
Neutrino Magnetic \vspace{0.5cm}Moment}\\
{\underline{Jo\~{a}o Pulido}
\footnote{
Talk presented at the XVI Autumn School CPMASS 97,
Lisbon, Portugal, 6-15 October 1997 (to appear in the Proceedings).
Work supported by CFIF.}
\& Ana M.\vspace{0.5cm} Mour\~{a}o
\footnote{This work was in part supported by GTAE-Portugal and JNICT projects 
PRAXIS/\-PCEX/\-P/\-FIS\-/4/\-96 and ESO/P/PRO/1127/96.}\\
{\small \sc CFIF $\&$ CENTRA, 
Instituto Superior T\'{e}cnico \\
Av. Rovisco Pais, 1096 Lisboa Codex, \vspace{0.5 cm}Portugal}

{ \small 
An analysis of neutrino electron scattering as applied to the SuperKamiokande
solar neutrino experiment with the data from the Homestake experiment leads to
an upper bound on the neutrino magnetic moment in the range 
$\mu_{\nu_e}\leq(2.9-3.7)\times 10^{-10}\mu_{B}$. This range is determined by
the spread in the flux predictions from six different standard solar models.
We assume equal magnetic moments for all neutrino flavours.
This limit is obtained when neutrinos do not undergo any "disappearance"
mechanism other than the magnetic moment conversion due to the solar magnetic
field and for a total or nearly total suppression of the intermediate energy 
neutrinos. We also point out that the limit may be further reduced if the 
threshold energy of the SuperKamiokande detector is decreased.}

\vspace { 15mm}

{\bf \large 1. Introduction}

\vspace { 6mm}

The solar neutrino problem, which first appeared as a deficit of the solar
neutrino flux in the Homestake experiment \cite{Davis88} relative to the 
solar model prediction \cite{Bahcall82}, has remained with us since its  
first acknowledgement in the late 1960's. In more recent years the Kamiokande
\cite{Hirata}, SAGE \cite{Abdurashitov}, Gallex \cite{Anselmann} and 
SuperKamiokande \cite{SuperKamiokande}
experiments, observing different parts of the neutrino spectrum, started 
operation. Besides these experiments, several theoretical solar models 
\cite{BP95} - \cite{FRANEC96} have been developed and our understanding of the
situation has changed. It now appears that the solar neutrino problem is not 
merely a deficit of the measured flux in the Kamiokande or the Homestake 
experiment. If it were so, it could be substantially reduced and even 
absorbed within the theoretical uncertainties in the $^{8}B$ neutrino flux
\cite{Crespo}, the only component observed in Kamiokande and the main one
in Homestake. More important, it is the problem of the disappearance of
the intermediate energy neutrinos \cite{Bahcall94} - \cite{B94}.
This is practically independent of any solar model considerations and 
relies essentially on a detailed analysis of the experimental data on the 
basis of the pp cycle dominance. There are therefore increasingly stronger
indications that the solution to the solar neutrino problem must rely on
non-standard neutrino properties, either neutrino oscillations in matter
\cite{W78}, vacuum \cite{H96}, the magnetic moment \cite{VVO,RSFP} or a
hybrid scenario \cite{MN}.

We will present here a new upper bound on the electron 
neutrino magnetic moment. Our work starts with an analysis on the dependence
of the neutrino survival probability on its energy and uses the most recent
data from the Homestake (Chlorine) and SuperKamiokande 
experiments. The first of these is 
looking at a purely weak charged current process, namely
\begin {equation}
\nu_{e}+^{37}{Cl}\rightarrow^{37}{Ar}+e^{-}
\end {equation}
whereas the second is based on elastic scattering,
\begin {equation}
\nu_{e,x}+e^{-}\rightarrow\nu_{e,x}+e^{-}
\end {equation}
with $x=\mu,\tau$  and
where possible electromagnetic properties of the neutrino may play a 
significant role. These are parametrised in terms of the electromagnetic form 
factors which at $q^2\simeq 0$ amount to the magnetic moment and charge 
radius. We allow for the solar neutrino deficit to be jointly explained in 
terms of these electromagnetic effects and any other sources like, for 
instance, oscillations. The upper bound on the magnetic moment is of course 
obtained when these other sources are absent. Previous analyses 
of solar neutrino data aimed at 
deriving bounds on the neutrino magnetic moment $\mu_{\nu}$ using neutrino 
electron scattering cross sections with electromagnetic interactions exist 
already in the literature \cite{Suzuki,MPR}. They did not however include the 
possibility of origins of the solar neutrino deficit other than the magnetic
moment transition, resulting therefore in upper and lower bounds for 
$\mu_{\nu}$. Furthermore they assumed an energy independent neutrino deficit,
which now appears not to be the case \cite{Bahcall94} - \cite{B94},
\cite{KP97}. 

Our results are derived for six
different theoretical solar models \cite{BP95} - \cite{FRANEC96}. 
They show a smooth dependence on $P_{I}$, the survival probability of the 
intermediate energy neutrinos, a parameter which to a very good accuracy
(better than $2\sigma$) can be assumed zero \cite{HR96,C97}. For all 
models we obtain an upper bound in the range $(2.9-3.7)\times {10^{-10}}
{\mu_B}$, an improvement
with respect to the most stringent laboratory bound existing to date,
$\mu_{\nu_e} \leq {6.1\times {10^{-10}} {\mu_B}}$ (90\% CL), from
the LAMPF group \cite{RPP}. More stringent bounds exist, however, for the
electron anti-neutrino magnetic moment at the same order of magnitude of
the numbers obtained here: $\mu_{\bar{\nu}_e} \leq {1.8\times {10^{-10}}
{\mu_B}}$ \cite{Derbin}. 

We restrict ourselves to the case of Dirac neutrinos. For Majorana neutrinos 
the analysis
would be different because an active $\bar{\nu}^{\rm M}_ {eR}$ could also 
be present and be detected through the process 
$\bar{\nu}^{\rm M}_ {eR}~ + p\rightarrow n+e^+$ for which there exists
however the firm upper bound from the Kamiokande II experiment  
$\Phi(\bar{\nu}^{\rm M}_ {eR})\leq (0.05-0.07)\Phi_{\nu_e}(^8B)$ \cite{santinu}.
Furthermore the states  $\bar{\nu}^{\rm M}_{\mu,\tau R}$
would now be active under weak interactions. 

The plan of the present work is to first derive in section 2 the possible
constraints of the survival probabilities of the intermediate and high energy
solar neutrinos which follow from the experimental data. In section 3 the 
expression for the event rate in the SuperKamiokande experiment is written in 
terms of these survival probabilities, the magnetic moment $\mu_{\nu}$ and 
the mean square radius $<r^2>$. From the lower laboratory bound on $<r^2>$
\cite{RPP} and the probability constraints, the upper bound on $\mu_{\nu}$
will follow. Finally in section 4 we derive our main conclusions and comment 
on possible future directions.

\vspace{10 mm}

{\bf \large  2. Energy Dependent Solar Neutrino Suppression}

\vspace{6 mm}
All six solar models (\cite{BP95} - \cite{FRANEC96}) whose relevant 
predictions are given in table I include heavy element 
diffusion except for TCL \cite{TCL} and TCCCD \cite{TCCCD}. 
It is now generally acknowledged that a 'standard'
solar model (SSM) should include diffusion, owing to the fact that such models 
give a remarkably good agreement with data from helioseismology \cite {SDI}.

The absence of the intermediate energy neutrinos, consisting principally of 
the $^7Be$ line at $E=0.86MeV$ and the CNO continuum, has been realised 
several years ago \cite{Bahcall94} from the comparison of the Homestake and
Kamiokande data. It is considered by some as the 'true' solar neutrino problem
in the sense that it is independent from normalisation to any solar model, 
either standard or non-standard \cite{DS96}. It appears as a natural 
consequence of the luminosity constraint \cite{BP95}, \cite{C97} 
($L_{\odot}=1.367\times10^{-1}Wcm^{-2}$)

\be
L_{\odot}=\sum_{k}(\frac{Q}{2}-\!<\!E_{\nu}\!>_k\!)\phi_{k}~~~(k=pp,pep,
^7\!\!{Be},CNO,^8\!\!{B})
\ee
with $Q=26.73MeV$ (total energy released in each neutrino pair production) and
the equations \cite {C97}
\bea
S_{Ga}&=&\sum_{i} \sigma_{Ga,i}\phi_{i}~~~(i=pp,pep,^7\!\!{Be},CNO,^8\!\!{B})\\
S_{Cl}&=&\sum_{j}\sigma_{Cl,j}\phi_{j}~~~(i=^7\!\!{Be},CNO,^8\!\!{B})\\
\phi_{pep}&=&0.021\phi_{pp}
\eea
where we used the weighted average from SAGE \cite{Abdurashitov} and Gallex 
\cite{Anselmann} ${\bar S_{Ga}}=73.8\pm 7.7SNU$ 
and the Chlorine data, $2.54\pm0.14\pm0.14SNU$ \cite{Homestake}.

One has in fact from this system of four equations, upon elimination of the 
$pp$ flux $\phi_{pp}$ and using the nuclear cross sections \cite {C97} 
$\sigma_{Ga,i}, \sigma_{Cl,j}$ the following intermediate energy neutrino 
flux (in units $cm^{-2}s^{-1})$
\bea
\phi_{Be}&=&1.04\times10^{4}\phi_{B}-2.88\times10^{10}\\
\phi_{CNO}&=&-8.46\times10^{3}\phi_{B}+2.22\times10^{10}.
\eea
Inserting $\phi_{B}$ from SuperKamiokande, \cite{SuperKamiokande} the total flux 
from these neutrinos is negative:
\bea
\phi_{Be}&=&-3.42\times10^{10}cm^{-2}s^{-1}\\
\phi_{CNO}&=&1.56\times10^{10}cm^{-2}s^{-1}.
\eea
Better fits were done by the authors of \cite{HR96},\cite{C97} who obtained
\bea
\phi_{Be+CNO}&\leq&0.7\times10^{9}cm^{-2}s^{-1}~~~~~(3\sigma)\\
\phi_{Be+CNO}&=&(-2.5\pm1.1)\times10^{9}cm^{-2}s^{-1}
\eea
which, compared with the theoretical predictions for six solar models
\cite{BP95} - \cite{FRANEC96} (see table I), gives
\be
P_I(3\sigma, \rm {all~six~models})\leq 0.16.
\ee

These authors used the former Kamiokande flux data which were higher than
Super\-Kamio\-kande. From equations (7) and (8) it is seen that the total flux
$\phi_{Be+CNO}$ decreases with decreasing $\phi_{B}$, so that the results 
(11), (12) should be further aggravated in the non-physical direction. Hence 
the probability that neutrinos are standard is no greater than 1\%, while, if 
the luminosity constraint is dropped, it may increase to 4\% \cite {HR96}. So 
intermediate energy neutrinos appear in practice to be completely suppressed.
\begin{center}
\begin{tabular}{|c|c|c|c|c|c|c|} \hline
Model&$R_{_{Cl}}^{_I}$&$R_{_{Cl}}^{_H}$&$R_{_{Cl}}$&$\phi_{_B}$&$R_{_{SK}}$&
$\phi_{Be+CNO}$ \\ \hline
BP95 \cite{BP95}&0.209&0.791&0.274&6.62&0.379&6.31 \\ 
TCL \cite{TCL}&0.248&0.752&0.401&4.43&0.567&5.37 \\ 
TCCCD \cite{TCCCD}&0.292&0.706&0.443&3.8&0.661&4.94 \\ 
P94 \cite{P94}&0.214&0.790&0.280&6.48&0.387&6.38 \\
RVCD96 \cite{RVCD96}&0.204&0.799&0.290&6.33&0.397&5.84 \\
FRANEC96 \cite{FRANEC96}&0.230&0.774&0.345&5.16&0.486&5.47 \\ \hline
\end{tabular}
\end{center}
Table I - { \small \sl The columns $R_{_{Cl}}^{_I}, R_{_{Cl}}^{_H}, R_{_{Cl}},\phi_{_B},
R_{_{SK}}, \phi_{Be+CNO}$ denote respectively the fractions of 
intermediate and high energy neutrinos in the Chlorine experiment, the ratio 
of the total measured signal and the model prediction, 
the $^{8}B$ flux prediction, the ratio data/model prediction for the 
SuperKamiokande data and the intermediate neutrino flux in each 
of the six models \cite{BP95} - \cite{FRANEC96}. Units of $\phi_{_B}$ are in
$10^6 cm^{-2}s^{-1}$ and units of $\phi_{BE+CNO}$ are in $10^9 cm^{-2}s^{-1}$.
}
\bigskip

As far as high energy ($^8B$) neutrinos are concerned and denoting by $R_{Cl}$
the ratio data/SSM prediction one may write
\be
R_{Cl}=R^{I}_{Cl}P_{I}+R^{H}_{Cl}P_{H}.
\ee
Here $R^{I(H)}_{Cl}$ is the fraction of intermediate (high) energy neutrinos 
in the Chlorine experiment as theoretically predicted and $P_{I(H)}$ is the
fraction of intermediate (high) energy $\nu_{e_L}$ produced in the Sun that
are detected on Earth. Using $P_{I}=0$ (99\%CL) and the models listed in 
table I one gets for $P_{H}$ the range
\be
0.35<P_H<0.63
\ee
with the smaller value corresponding to BP95 \cite{BP95} and the larger to 
TCCCD \cite{TCCCD}.

This will be the range of values used for $P_H$ in the following section.

\vspace{10 mm}

{\bf \large  3. Event Rates and Cross Sections}

\vspace{6 mm}

The event rate in a solar neutrino experiment in which recoil electrons are 
produced is given by the corresponding cross section per unit neutrino energy 
$E_{\nu}$ per unit kinetic energy $T$ of the recoil electron times the neutrino
flux and summed over all possible neutrino fluxes:
\begin{equation}
S_{exp}=\sum_{i} \int dE_{\nu_i} \int \frac{d^{2}\sigma}{dT dE_{\nu_i}} 
\phi(E_{\nu_i}) dT \label{sexp}
\end{equation}
The quantity $\phi(E_{\nu_i})$ represents the i-th normalised neutrino flux. 
For SuperKamiokande, which is based on neutrino electron scattering, and where 
only the $^{8}B$ neutrino flux is seen, we have
\begin{equation}
S_{SK}=\!\int dE_{\nu}\int \phi(E_{\nu})\left( X_W\frac{d^2\sigma_W}{dT 
dE_{\nu}}+\frac{d^2\sigma_{+EM}}{dT dE_{\nu}}+\frac{d^2\sigma_{-EM}}{dT 
dE_{\nu}}+X_{int}\frac{d^2\sigma_{int}}{dT dE_{\nu}}\right) dT
\end{equation}
The quantities $X_W$, $X_{int}$ will be derived below. The weak 
($d^2\sigma_W/dTdE_{_\nu}$), electromagnetic spin non-flip 
($d^2\sigma_{+EM}/dTdE_{_\nu}$), 
electromagnetic spin flip ($d^2\sigma_{-EM}/dTdE_{_\nu}$) and interference
($d^2\sigma
_{int}/dTdE_{_\nu}$) parts of the differential cross section were taken 
from \cite{KSN}.
Denoting by $\phi_{\nu}$ the neutrino magnetic moment in Bohr magnetons $\mu_B$ we 
have 
\bea
\frac{d^2\sigma_W}{dT dE_{\nu}}&\!=\!&\!\!\!\frac{G^2_{F}m_{e}}{2\pi}
\left((\gv  +\ga )^2+
(\gv -\ga )^2(1-\frac{T}{E_{\nu}})^2-(\gv ^2-\ga ^2)\frac{m_{e}T}{E_{\nu}^2}
\right) \\
~\nonumber\\
\frac{d^2\sigma_{+EM}}{dT dE_{\nu}}&\!=\!&\!<r^2>^2\frac{\pi\alpha^2}{9}m_e
\left( 1+(1-\frac{T}{E_{\nu}})^2-\frac{m_{e}T}{E_{\nu}^2}\right) \\
~\nonumber\\
\frac{d^2\sigma_{-EM}}{dT dE_{\nu}}&\!=\!&\!f_{\nu}^2\frac{\pi\alpha^2}{m_e^2}
\left( \frac{1}{T}-\frac{1}{E_{\nu}}\right) \\
~\nonumber\\
\frac{d^2\sigma_{int}}{dT dE_{\nu}}&\!=\!&\!\!\!-<r^2>\frac{\sqrt{2}}{3}
\alpha G_{F}m_e
\left( \gv \frac{m_{e}T}{E_{\nu}^2}-(\gv +\ga )-(\gv -\ga )(1-
\frac{T}{E_{\nu}})^2\right)~~,\nonumber  \\
~~
\eea
where
$
\gv=-1/2+2~sin^2\theta_W$, ~$\ga=-1/2$
for $\nu=\nu_{\mu},\nu_{\tau}$ and 
 $~\gv=1/2+2~sin^2\theta_W,~\ga=1/2$
for $\nu=\nu_e$. We use $sin^2\theta_W=0.23$.
There are upper and lower experimental bounds for the mean square radius of 
the neutrino \cite{RPP} (90\% CL):
\be
-7.06\times10^{-11} <~ <r^2>~ < 1.26\times10^{-10} MeV^2.
\ee
From the inequality \cite{MPR}
\be
E_{\nu}\geq\frac{T+\sqrt{T^2+2m_{e}T}}{2}~~,
\ee
the maximum $^8$B neutrino energy \cite{Bahcall82} $E_{\nu_M}=15MeV$
and the electron threshold energy in the SuperKamiokande detector
$E_{e_{th}}$=7 MeV, one can derive the lower and upper integration limits
in eq. (17).

It should be noted at this stage that the integrated cross section in (17) 
refers to a neutrino flux which is assumed to have been modified either due
to the magnetic moment spin flip inside the Sun or through flavour 
oscillations in the Sun or on its way to the detector. So an electron neutrino 
from the $^{8}B$ flux produced in the core of the Sun has a survival 
probability $\ph $ of reaching the SuperKamiokande detector, thus interacting 
weakly with the electron via the neutral or the charged current. The remaining
$(1-\ph)$ fraction of the flux will have oscillated to $\nu_{\mu L}$ (or 
$\nu_{\tau L}$) with a probability $\alpha$, thus interacting via the weak 
neutral
and electromagnetic currents only. Alternatively it will have 
flipped to $\nu_{e R}$ 
(or ${\nu}_{{\mu,\tau} R}$) with a probability $(1-\alpha)$ via the magnetic 
moment, thus interacting only through the electromagnetic current (see fig.1). 
\begin{figure}
\begin{picture}(6,6)
\put(4,-1){\psfig{figure=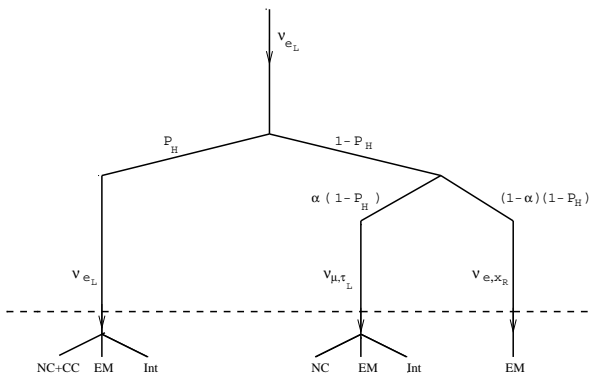,angle=270,width=8cm}}
\end{picture}
\caption{\small \sl  A fraction $P_{H}$ of the initial $\nu_{e_L}$ flux 
remains unaltered and interacts with $e^-$ in SuperKamiokande. 
 Its cross section contains a weak 
contribution (charged (CC) and neutral current (NC)), an electromagnetic one
and the interference between them. Of the remaining $(1-P_H)$, a fraction
$\alpha$ is converted to $\nu_{\mu,\tau L}$ and interacts without the weak 
charged current while the remaining $(1-\alpha)(1-P_H)$ interacts only
electromagnetically.}
\end{figure}

\bigskip

The weak part of the total cross section in SuperKamiokande $\sigma^K_W$ may 
therefore be decomposed as follows
\bea
\sigma^K_W&=&\ph \sigma_{_W}+\alpha (1-\ph)
\sigma_{NC}\nonumber \\
&\simeq& \sigma_{_W}(0.15\alpha+\ph(1-0.15\alpha))
\eea
where $\sigma_{NC}$ denotes the weak neutral cross section and $\sigma_{_W}$ 
denotes the total $\nu_{e}e$ cross section which includes the neutral and
charged current contributions.
In eq. (24) we have used the well known fact that \cite{Okun}
\be
\sigma_{_W}\simeq6.7\sigma_{_{NC}}.
\ee
This yields the parameter $X_W$ in equation (17):
\be
X_W=0.15\alpha+\ph(1-0.15\alpha).
\ee
In order to determine $X_{int}$, we decompose the interference cross section
[eq.(21)] into its $\nu_e$ and $\nu_{\mu,\tau}$ parts, recalling as above 
that $\nu_{eL}$ has 
partly survived with probability $\ph $ and partly oscillated to 
$\nu_{\mu,\tau L}$
with probability $\alpha(1-\ph)$:
\bea
\sigma_{int}^K &= &\ph \sigma_{{\nu_e},int}+\alpha(1-\ph)
\sigma_{{\nu_{\mu},int}} \nonumber \\
&\simeq &\sigma_{{\nu_e},int}(\ph-0.37~\alpha~(1-\ph)).
\eea
In the last step we used (21) and the definitions above for $\ga,\gv$ to obtain
\be
\frac{\sigma_{\nu_{\mu,int}}}{\sigma_{\nu_{e,int}}}\simeq-0.37
\ee
for the integrated cross sections, which yields \footnote{Since we are
interested in the upper bound for the magnetic moment which is obtained as
will be seen for vanishing charge radius, we assume 
$<r^2>_{\nu_e}=<r^2>_{\nu_{\mu,\tau}}$ and $\mu_{\nu_e}=\mu_{\nu_{\mu,\tau}}$}
\be
X_{int}=(\ph-0.37~\alpha~(1-\ph)).
\ee

If neutrinos are standard, they do not oscillate nor have any electromagnetic
properties and only the $\sigma_W$ term survives in equation (17). This 
corresponds to $X_W=1$
($\alpha=0,\ph=1$). In such a case the prediction of eq. (17) for the 
SuperKamiokande event rate is wrong by a solar model dependent factor $R_K$ 
which is the ratio between the data and the model prediction:
\be
S_{SK}=R_{SK}\int dE_{\nu}\int \phi(E_{\nu})\frac{d^2\sigma_W}{dT dE_{\nu}}dT.
\ee
The basic point of the paper is to equate the right hand sides of (17) and 
(30). We note that in doing so we are not merely attempting to explain the
neutrino deficit in SuperKamiokande which is model dependent. Even if 
$R_{SK}=1$ (no neutrino deficit appears in SuperKamiokande) there may still be 
electromagnetic properties related to the main problem of the disappearance 
of the intermediate energy neutrinos.

Equating (17) and (30) and taking $R_{SK}$ as an input, leaves us four 
parameters 
($\alpha,\ph$ and the electromagnetic ones -- $f_{\nu}$, $<r^2>$) of which 
$\ph$ is directly related to $P_{_I}$ (see eq.(14)). We obtain
\bea
f_{\nu}^2\!\!&=&\!\!\!\! \left( R_{SK}-0.15~\alpha-\ph(1-0.15~\alpha~) \right)
                   \frac{\sigma_W}{B_{-EM}} \nonumber\\
     &-&\!\!<\!r^2\!>\! \left( \ph-(1-\ph)~0.37~ \alpha~ \right)
          \frac{A_{int}}{B_{-EM}}-
       \!<\!r^2\!>^2\!\frac{B_{+EM}}{B_{-EM}}
\eea
where 
\bea
\sigma_W&=&\int dE_{\nu} \int \phi(E_{\nu})\frac{d^2\sigma_W}{dT dE_{\nu}}dT\\
<r^2>^2B_{+EM}&=&\int dE_{\nu} \int \phi(E_{\nu})
\frac{d^2\sigma_{+EM}}{dT dE_{\nu}}dT \\
f_{\nu}^2 B_{-EM}&=&\int dE_{\nu}\int \phi(E_{\nu})\frac{d^2\sigma_{-EM}}{dT 
dE_{\nu}}dT\\
<r^2>A_{int}&=&\int dE_{\nu}\int \phi(E_{\nu})\frac{d^2\sigma_{int}}{dT 
dE_{\nu}}dT.\\
~~\nonumber
\eea

For a given $R_{SK}$, maximising the magnetic moment for fixed $P_H$ amounts 
to minimising $\alpha$ and $<r^2>$ ((see equation (31)). 
This is to be expected since it corresponds to the absence of oscillations and 
minimal mean square radius. 

Refering to the six solar models above \cite{BP95} - \cite{FRANEC96} and
using \cite{SuperKamiokande} $\phi_{^8_B}=(2.51\pm^{0.14}_{0.13}\pm 0.18)
\times 10^6 cm^{-2}s^{-1}$ with a threshold $ E_{e_{th}}=7.0MeV$, we
display in figs. 2, 3 the magnetic moment 
$\mu_{\nu_{e}}$ as a function of $<r^2>$ in the limit $\alpha=0$ and as a 
function of $\alpha$ in the limit $<r^2>=<r^2>_{min}$ respectively.

\begin{figure}[ht]
\begin{picture}(6,6)
\put(4,-0.3){\psfig{figure=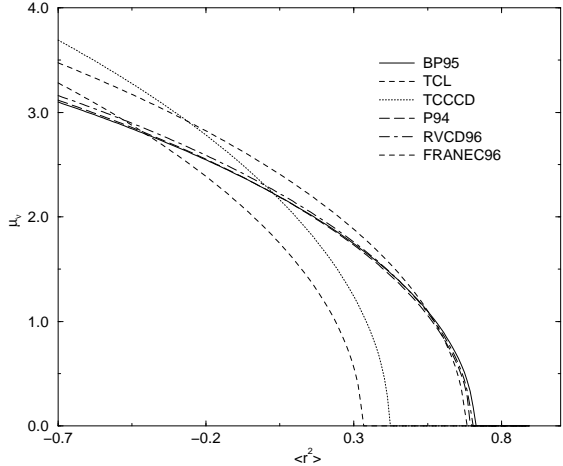,angle=270,width=9cm}}
\end{picture}
\caption{\small \sl Neutrino magnetic moment (in units $10^{-10}\mu_B$)
as a function of the mean square radius
$<r^2>$ (in units $10^{-10}MeV^{-2}$) in the limit ${\alpha}=0$
and in each of the six models [7]-[12].
The upper bound on $\mu_{\nu_e}$ is in each model the left end
of the curve.}
\end{figure}

\begin{figure}
\begin{picture}(6,6)
\put(4,-0.3){\psfig{figure=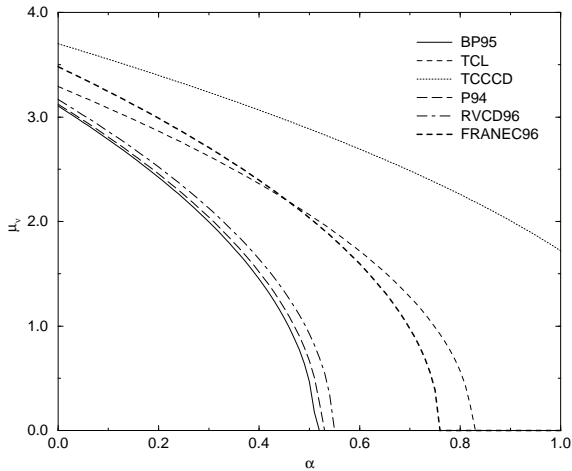,angle=270,width=9cm}}
\end{picture}
\caption{\small \sl Same as {\rm fig.2} as a function of $\alpha$ in the limit 
$<r^2>=<r^2>_{min}$.}  
\end{figure}

As shown above, up to more than $2\sigma$ one can take $P_{_I}=0$, so it is 
appropriate to consider the left ends of these curves as the actual upper 
limits on $\mu_{\nu_e}$ from experiment and theoretical models. We have in 
these conditions
\be
\mu_{\nu_e}\leq(2.9 - 3.7)\times 10^{-10}\mu_{_B}~~\\
\ee
We also note, as can be seen from figs. 2, 3,
that the disparities on the predictions for the $^8{B}$ flux among 
solar models (table I), related to uncertainties in the astrophysical factor
$S_{17}$, are hardly reflected on the upper bound on $\mu_{\nu}$ for all 
neutrino types.

An essential development which may further improve the bound (36) is the 
decrease in $E_{e_{th}}$, the recoil electron threshold energy in 
$\nu_{_{e,x}}e$ scattering. This decrease implies a
decrease in the ratio of integrals $\sigma_{_W}/B{_{-EM}}$ appearing in
equation (31). This is related to the fact that for decreasing energy and a
sizable neutrino magnetic moment, the electromagnetic contribution to the
scattering increases faster than the weak one. The above referred ratio of
integrals leads through (36) and for constant values of $R_{SK}$ and $P_{_H}$
to a decrease in the upper bound for $f_{\nu}$. 
The SuperKamiokande collaboration so far has operated with a threshold of 
7.0 MeV and plans to improve it down to 
5.0 MeV in the near future. The forthcoming SNO experiment \cite{SNO} also
aims to operate near this threshold. For $E_{e_{th}}$ =5.0 MeV
and the same ratio of data/model prediction for the
$^8{B}$ neutrino flux ($R_{SK}$), the bound (36) would be 
decreased by approximately 50\%. Hence a further decrease 
in the electron threshold energy will be a welcome improvement.

\vspace {10 mm}

{\bf \large 4. Conclusions}

\vspace {6 mm}

We have investigated the existence of an upper bound on the electron neutrino
magnetic moment $\mu_{\nu_e}$ from solar neutrino experiments. Besides 
laboratory bounds, this looks a promising source for constraining 
all neutrino magnetic moments and thus establishing upper limits 
on these quantities. The 
strictest laboratory bounds existent up to date refer to electron 
anti-neutrinos ($\mu_{\bar \nu_e}<1.8\times 10^{-10}\mu_{_B}$ 
\cite{Derbin}) and a new experiment \cite{NIM} 
aimed at providing new constraints is expected to start operation soon. 
Regarding laboratory bounds on $\mu_{\nu_e}$, the limit is higher: 
$\mu_{\nu_e}<6.1\times 10^{-10}\mu_{_B}$ \cite{RPP}. We believe the 
present work, where we used SuperKamiokande data, improves this bound 
by a factor of approximately 2. 
We find $\mu_{\nu_e}<(2.9-3.7)\times 10^{-10}\mu_{_B}$. Both were obtained on
the assumption of equal neutrino magnetic moments for different flavours.
Furthermore we assumed a total suppression of intermediate energy neutrinos: 
$P_{_I}=0$.

From the solar models standpoint, the uncertainties in $S_{17}$, the parameter
describing the $^8{B}$ flux prediction, although not irrelevant, do not play
a crucial role. In fact, the upper bound on $\mu_{\nu_e}$ is only very 
moderately sensitive to them.

On the other hand, the decrease in the recoil electron threshold energy in the
solar neutrino electron scattering may further constrain this bound. Thus not
only the expected improvement in SuperKamiokande, but also the SNO experiment
\cite{SNO} examining this process with a 5 MeV threshold or possibly lower 
will be essential for the purpose.

\end{document}